\documentclass[11pt]{article}
\pdfoutput=1
\usepackage{egame}
\usepackage{setspace} 
\usepackage{amsmath}
\usepackage{amssymb}
\usepackage{algorithm}
\usepackage{algorithmic}
\usepackage{color, graphics}
\usepackage[all]{xy}
\usepackage{setspace} 
\usepackage{rotating}
\usepackage{hhline}
\usepackage{multirow}
\usepackage{natbib}

\newcommand{\bs}{\boldsymbol}

\newcommand{\citeposs}[1]{\citeauthor{#1}'s \citeyearpar{#1}}

\title{A Statistical View of Learning in the Centipede Game}

\author{Anton H. Westveld$^a$ and Peter D. Hoff$^b$\\ \\
$^a$Assistant Professor of Statistics, Department of Mathematical Sciences,\\ University of Nevada, Las Vegas, NV \\e-mail: anton.westveld@unlv.edu\thanks{I  would especially like to thank Julian Besag.  The approach taken in this paper of using randomization tests and then comparing them to posterior predictive tests after modeling was his notion.   Additionally, I thank Grace S. Chiu,  Andrew D. Martin, Kevin M. Quinn, and Michael D. Ward for discussions related to this project and examining various iterations of this document.} \\\\ $^b$Associate Professor, Department of Statistics,\\ University of Washington, Seattle, WA\\e-mail: pdhoff@uw.edu}

\begin{document}
\bibliographystyle{apsr}

\maketitle

\begin{abstract}
In this article we evaluate the statistical evidence that a population of students learn about the sub-game perfect Nash equilibrium of the centipede game via repeated play of the game. This is done by formulating a model in which a player's error in assessing the utility of decisions changes as they gain experience with the game. We first estimate parameters in a statistical model where the probabilities of choices of the players are given by a Quantal Response Equilibrium (QRE) \citep{McKPal95,McKPal96,McKPal98}, but are allowed to change with repeated play. This model gives a better fit to the data than similar models previously considered. However, substantial correlation of outcomes of games having a common player suggests that a statistical model that captures within-subject correlation is more appropriate. Thus we then estimate parameters in a model which allows for within-player correlation of decisions and rates of learning.  Through out the paper we also consider and compare the use of randomization tests and posterior predictive tests in the context of exploratory and confirmatory data analyses.
\end{abstract}


\pagebreak
\section{Introduction}
Decision making under uncertainty has long been of interest to a wide variety of academic disciplines: biology, computer science, economics, mathematics, philosophy, political science, and statistics, to name a few.  The main mathematical method for examining multi-agent decision theory has been game theory. However, the game theoretic solutions of some simple games have been called into question, with a classic example being the Sub-Game Perfect Nash Equilibrium (SPNE) of the centipede game. In experimental settings, individuals rarely choose the SPNE solution \citep{McKPal93}.  To explain this, \citet{McKPal95,McKPal96,McKPal98} suggest that players' strategies can be represented by a Quantal Response Equilibrium (QRE), in which players' choices deviate from the SPNE because of ``mistakes'' in decision making.  The mistakes or errors may be due to lack of information, information overload, even the fact that human beings are not perfect optimizers, or as is often the case they are not optimizing according the specific criterion set out by researchers in a given study.

Our examination of the data collected by \cite{McKPal93} on the centipede game shows that on average players move toward the SPNE with repeated play.  This idea of moving toward a game theoretic equilibrium through repeated play has been called \textit{learning} by \citet{FunTir91}.  An extensive amount of theoretical work has been written on the subject, as well as a fair amount of empirical work based on the centipede game \citep{ElgMcKPal93}.  We expand the QRE framework, which allows for a statistical interpretation of game theoretic models, by allowing the error distribution to change as players gain experience.  We also build upon the notion of heterogeneity of players, discussed by \cite{McKPal96} through the introduction of different parameters for each type of player in the game and finally expanding that notion to a statistical random effects model that allows for heterogeneity over all the subjects in the data set. The models we employed represent the data better than previous models based on the Bayesian Information Criteria (BIC) as a measure of adequacy \citep{KasRaf95}.  The outline of the paper is as follows: In Section 2, the game and the experimental design are discussed.  In Section 3, an exploratory data analysis is presented.  In Section 4, several models are examined that allow the distribution of a player's error to change through experience.  In Section 5, the model with the best BIC is developed further to allow for heterogeneity of players in the data through a random effects model which accounts for the correlation of outcomes involving a common player. The paper then ends with a discussion of model limitations and the potential for future investigation.

\section{The Game and Experimental Design}
\begin{figure}[htb]
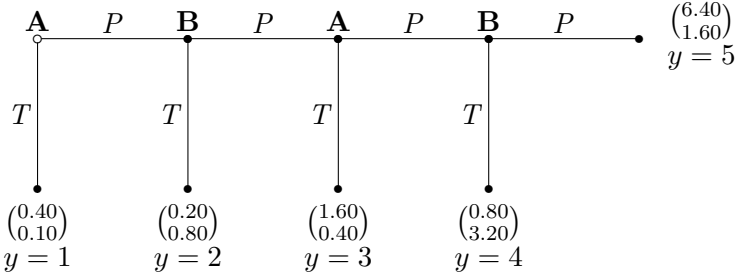

\hspace*{\fill}
\begin{egame}(700,700)
    \putbranch(0,350)(0,1){200}
    \iib{\textbf{A}}{\textit{T}}{\textit{P}}
    [$\begin{array}{c} \binom{0.40}{0.10}\\y=1 \end{array}$][]

    \putbranch(200,350)(0,1){200}
        \iib{\textbf{B}}{\textit{T}}{\textit{P}}
    [$\begin{array}{c} \binom{0.20}{0.80}\\y=2 \end{array}$][]

    \putbranch(400,350)(0,1){200}
        \iib{\textbf{A}}{\textit{T}}{\textit{P}}
    [$\begin{array}{c} \binom{1.60}{0.40}\\y=3 \end{array}$][]

    \putbranch(600,350)(0,1){200}
        \iib{\textbf{B}}{\textit{T}}{\textit{P}}
    [$\begin{array}{c} \binom{0.80}{3.20}\\y=4 \end{array}$][$\begin{array}{c} \binom{6.40}{1.60}\\y=5 \end{array}$]
\end{egame}
\hspace*{\fill} \caption[]{M-P four-stage centipede game; T = Take, P = Pass; $y$ denotes the outcome of the game.}
\label{gt4low}
\end{figure}

The data were gathered by \cite{McKPal93} based upon the four-stage centipede game shown in Figure \ref{gt4low}.  A single run of the centipede game involves two player types --- Player A and Player B.  Player A initiates the game and in the first stage has an opportunity to either Take or Pass.  If Player A chooses Take, the game ends and Players A and B receive 40 and 10 cents, respectively.  If Player A passes, then Player B has an opportunity to either choose Take or Pass. Again, if Player B chooses Take the game ends and Players A and B receive 20 and 80 cents, respectively.  At each subsequent stage the dollar amounts are doubled and switched between Player A and Player B.  The fourth stage is the last regardless of whether Player B chooses Take or Pass.  Based upon this pattern, there are five possible outcomes of the game ($y =\{1,\ldots,5\}$), where an outcome is the total number of stages played in the game. The traditional game theoretic solution, the SPNE, can be determined via backwards induction; at the fourth stage, based upon utility maximization under the assumption that a Player's utility is determined solely by the monetary outcomes, it seems natural for Player B to choose Take.  If the game were to reach stage 3, then Player A should realize this and following a similar argument would choose Take in the third stage. This continues backwards through the game tree, yielding the unique solution that Player A should choose Take at the first stage. However, this solution is Pareto inferior \citep{MasWhiGre95}, since both players would strictly benefit by moving further out in the game (stage 3 and beyond).

\vspace{-1.2in}

The data were collected in three different sessions, two of which consisted of 20 and 18 students from Pasadena Community College, and a third session of 20 students from the California Institute of Technology.  Each subject took part in only one of the three sessions.  In each session, subjects were randomly assigned to be either a Player A or Player B and this assignment was kept throughout their allotted session.  In the first and third sessions, subjects played 10 games, while in the second session they played only 9 games. After each game, the subjects who were Players B were rotated so that two subjects never played each other more than once.  The layout for the two 10-game experiments can be seen in Table \ref{LSD}.  The subjects who are Players A and Players B are on the rows and columns, respectively.  The $(i,j)^{th}$ entry of the table indicates the game number played between the $i^{th}$ Player A and the $j^{th}$ Player B.  Since each individual only plays 10 games the experimental design is a Latin square (i.e. in each row and column each game number appears only once).  Notice that we could place the game number on the columns and fill in the table the Players B and we still would have a Latin square. In fact any permutation of Players A, Players B, and game number with rows, columns and table entries is a Latin square.  Let $y_{[i,j](s)}$ be the outcome $\{1, \ldots, 5\}$ for the game played between the $i^{th}$ Player A and $j^{th}$ Player B in the $s^{th}$ session, where  $i =\{1, \ldots, N(s)\}$ and $j =\{1, \ldots, N(s)\}$.  Since session 2 has only 18 subjects $N(s=2)=9$, compared to $N(s=1,3)=10$  Thus the total number of cases in the data are $\sum_{s=1}^3 N(s)\times N(s) = 10^2+9^2+10^2 = 281$. 

\begin{table}[!h]
\begin{footnotesize}
\begin{center}
\begin{tabular}{rr|rrrrrrrrrr}
    \multicolumn{12}{c}{Player B} \\ \\ 
\multirow{11}{*}{\begin{sideways} Player A \end{sideways}} & & $B_1$ & $B_2$ & $B_3$ & $B_4$ & $B_5$ & $B_6$ & $B_7$ & $B_8$ & $B_9$ & $B_{10}$ \\
  \hline
 & $A_1$ & 1 & 2 & 3 & 4 & 5 & 6 & 7 & 8 & 9 & 10 \\
 & $A_2$ & 10 & 1 & 2 & 3 & 4 & 5 & 6 & 7 & 8 & 9 \\
 & $A_3$ & 9 & 10 & 1 & 2 & 3 & 4 & 5 & 6 & 7 & 8 \\
 & $A_4$ & 8 & 9 & 10 & 1 & 2 & 3 & 4 & 5 & 6 & 7 \\
 & $A_5$ & 7 & 8 & 9 & 10 & 1 & 2 & 3 & 4 & 5 & 6 \\
 & $A_6$ & 6 & 7 & 8 & 9 & 10 & 1 & 2 & 3 & 4 & 5 \\
 & $A_7$ & 5 & 6 & 7 & 8 & 9 & 10 & 1 & 2 & 3 & 4 \\
 & $A_8$ & 4 & 5 & 6 & 7 & 8 & 9 & 10 & 1 & 2 & 3 \\
 & $A_9$ & 3 & 4 & 5 & 6 & 7 & 8 & 9 & 10 & 1 & 2 \\
 & $A_{10}$ & 2 & 3 & 4 & 5 & 6 & 7 & 8 & 9 & 10 & 1 \\
\end{tabular}
\end{center}
\end{footnotesize}
\caption{Latin square design for twenty subjects (10 Players A and 10 Players B) --- the game number is a table entry.}
\label{LSD}
\end{table}

Finally, In an attempt to conform to the notions of rationality required by game theoretic solutions, the structure of the game, number of times the game would be played, and payment structure were made common knowledge to all the subjects.  This was done by reading a set of instructions, a practice session, as well as the administration and correction of a quiz.  It is important to note that the subjects were not ``taught'' what an optimal strategy was in any sense.  Additionally, the games were conducted on computers so that the subjects did not know whom they were playing against and at the end of each session, the subjects were privately paid the amount of money they had earned from the 9 or 10 games.  Further discussion of the experimental design and data collection can be found in \cite{McKPal93}.

\section{Exploratory Data Analysis}
\begin{figure}[htb]
\begin{center}
{\scalebox{0.7}{\includegraphics{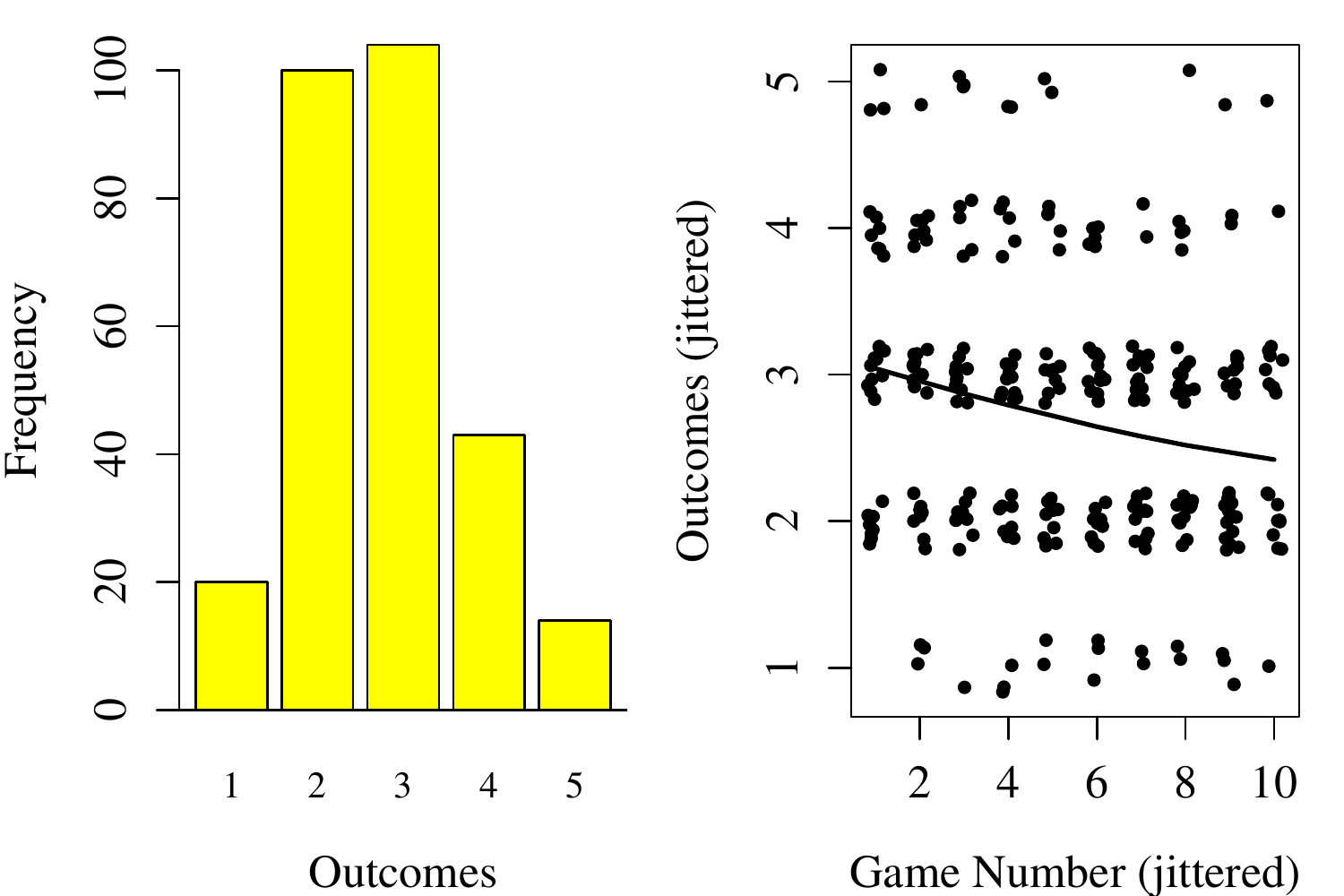}}} 
\caption{The histogram of the five outcomes and a scatter plot of the number of games played against the five outcomes with a loess smoother.}
\label{edaHistScatt}
\end{center}
\end{figure}

The left panel of Figure \ref{edaHistScatt} presents the frequencies of the outcomes for all the games played for the combined sessions.  The traditional game theoretic solution is for Player A to choose Take at the first stage. If the subjects actually played in this manner all the mass in the histogram would be contained on outcome 1.  However,  most of the mass occurs on outcomes 2 and 3.  Surprisingly, we see some mass on outcome 5 even though we would expect a subject reaching this stage to examine their payoffs  of \$3.20 versus \$1.60 and choose Take.

Since it is clear that most subjects do not play the SPNE, a primary scientific question of interest is whether subjects, through repeated gaming, move toward the SPNE.  In the right panel of Figure \ref{edaHistScatt}, a scatter plot of the number of games played against the five possible outcomes of the game is presented with a locally smooth regression \footnote{The default setting of the \texttt{lowess()} function in the R statistical package was used.}. Due to the discrete nature of the data the points were jittered.  The decreasing trend of the smoother suggests that on average the subjects move toward the first outcome with repeated play, which is the SPNE.  

\subsection{Randomization Test of Trend}
\begin{figure}[htb]
\begin{center}
{\scalebox{0.6}{\includegraphics{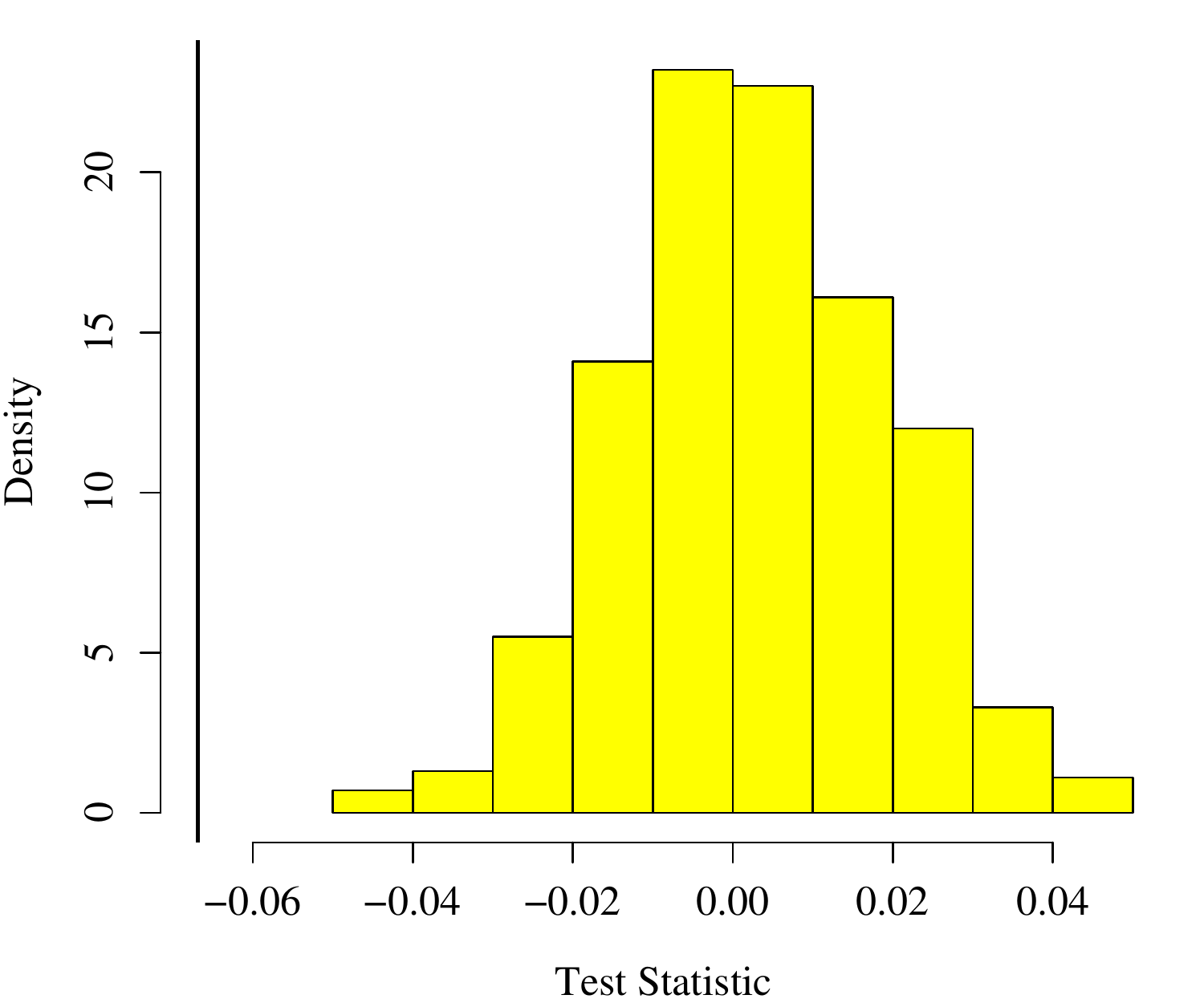}}} 
\caption{The approximate null distribution and observed test statistic are compared.  The slope of number of games played against the five outcomes was used as the test statistic of interest and the associated p-value for the randomization test was 0.}
\label{MCpv1}
\end{center}
\end{figure}

The smoother suggests that the relationship between the number of games played and the outcomes of the games is approximately linear.  The slope of the linear component of the trend was estimated as $\gamma_{obs}=-0.067$ via least squares.  Because of the potential for dependent outcomes due to repeated play by each subject, typical regression standard errors are inappropriate.  To overcome this, we conducted a randomization test to examine whether the observed slope was statistically different from zero \citep{Fis35,BoxAnd55,Bes77}.  As \citeposs{Bes77} state ``a primary advantage of [randomization] testing is that the investigator is free to use a variety of informative statistics of his own choosing, rather than be dictated to by known distributional theory.  Indeed, even when the relevant asymptotic distribution theory is available, [randomization] testing provides an exact alternative for small samples.''  Thus in the randomization testing framework, which does not consider tests based on population parameters, our null hypothesis is $H_0$: the game number does not affect the outcome.  In order to investigate this hypothesis, we consider the slope as our chosen test statistic.  The test proceeds by randomly sampling appropriate permutations of the data $\bs{y}_{perm}$ under $H_0$.  Then, for each permutation computing $\gamma(\bs{y}_{perm})$, and finally comparing this null distribution to the statistic calculated from the observed data $\gamma(\bs{y}_{obs})$.  We now provide further details on this procedure.  In conducting the test we need to be faithful to the design, so the permutations were done according to a Latin square design \citep{Cox58}.  For each of the three sessions, the data can be represented by a Latin square with the Players A represented on the rows and Players B on the columns as in Table \ref{LSD}.  For each pair of players $A_i$ and $B_j$, information exists on the outcome of the game that the pair played, as well as the current game number.  The rows and columns of this matrix were permuted while keeping fixed the row and columns labels, and the outcome of the game for that pair.  Each permutation shuffled the ``times''  at which the games were played but maintained who played each game and the outcome.  This was done for each session. The data from the three sessions were placed together and the slope of the linear trend was estimated  $\gamma(\bs{y}_{perm})$.  One thousand values of $\gamma(\bs{y}_{perm})$ were sampled in this manner, and compared to $\gamma(\bs{y}_{obs})$.  The results of the randomization test are displayed in Figure \ref{MCpv1}.  The approximate one sided p-value, $P[\gamma(\bs{y}_{perm}) \leq \gamma(\bs{y}_{obs})|H_0]$, was $0$ (i.e., none of the statistics from the randomization met or exceeded the observed value) suggesting that we reject the null hypothesis. 

\section{Multinomial Models}
Since the outcomes of the centipede game take on five values it is natural to model the data using a multinomial distribution --- the key question pertains to the parameterization of the probabilities of each outcome occurring.  The SPNE is perhaps the simplest model and states that probability of the first outcome is always one, $P[y_{[i,j](s)}=1]=1$ $\forall$ $i,j,s$, which is clearly not a good model for these data.  Considering this, McKelvey and Palfrey, in a series of papers  \citep{McKPal95,McKPal96,McKPal98} relaxed this criterion  through the development of the Quantal Response Equilibrium (QRE) model based on the work in statistical choice modeling by \citet{McFad73}.  Their model still uses the decision making process to inform the specification of the probabilities of the five outcomes, but differs from the SPNE by allowing players to make mistakes through a stochastic component added to players' decisions. We expand upon this model to capture the observed mean trend over time which could be interpreted as \textit{learning}.  We also expand the model to allow for heterogeneity in the type of players (i.e. whether the subject is a Player A or Player B). The section ends with a Bayesian analysis of the model with the best BIC and confirmatory data analysis that will be used to motivate the random effects model of the following section.

\subsection{McKelvey and Palfrey's Original Models}
\citet{McKPal98} fit two different QRE models to the four-stage centipede data as examples of the QRE methodology which they developed \citep{McKPal95,McKPal96,McKPal98}.  We will present their one-parameter QRE model since it will serve as the basis for the models presented in this paper.  The QRE model parameterizes the probabilities of players' decisions as functions of their payoffs and the precision (or variance) of their errors. Based upon the extensive form of the game depicted in Figure \ref{gt4low}, the decision probabilities that need to be specified for each pair $A_i, B_j$ within each session $s$ are:

\small
\begin{eqnarray*}
q2_{[i,j](s)} & = & P[\text{Player $B_j$ chooses Take at stage 4}|\text{Player $B_j$ reaches stage 4 when playing against $A_i$}],\\
p2_{[i,j](s)} & = & P[\text{Player $A_i$ chooses Take at stage 3}|\text{Player $A_i$ reaches stage 3 when playing against $B_j$}],\\
q1_{[i,j](s)} & = & P[\text{Player $B_j$ chooses Take at stage 2}|\text{Player $B_j$ reaches stage 2 when playing against $A_i$}],\\
p1_{[i,j](s)} & = & P[\text{Player $A_i$ chooses Take at stage 1 when playing against $B_j$}].\\
\label{eq2}
\end{eqnarray*}
\large

For example, for a Player B to choose Take at the fourth stage, the perceived utility gained from that choice should be greater than the perceived utility gained by choosing Pass. The perceived utilities that drive a Players B decision are modeled as $U_B^*(y_{[i,j](s)}=4) =U_B(y_{[i,j](s)}=4) + \alpha(q2)_{[i,j](s)}$ versus $U_B^*(y_{[i,j](s)}=5) =U_B(y_{[i,j](s)}=5) + \gamma(q2)_{[i,j](s)}$, where $U_B(y_{[i,j](s)}=4)$ and $U_B(y_{[i,j](s)}=5)$ are the monetary payoffs for the outcomes four and five.  For the rest of this paper we will make this assumption, however the authors note that with larger monetary values other utility functions may be more appropriate.  Finally, the $\alpha$'s and $\gamma$'s are random deviations that can vary across players, games, and stages within a single game.  Therefore the probability that a Player B chooses Take is:

\begin{eqnarray*}
q2_{[i,j](s)} &=& P[U_B(y_{[i,j](s)}=4) + \alpha(q2)_{[i,j](s)} > U_B(y_{[i,j](s)}=5) + \gamma(q2)_{[i,j](s)}]\\
&=& P[3.20 + \alpha(q2)_{[i,j](s)} > 1.60 +\gamma(q2)_{[i,j](s)}]\\
&=&P[\gamma(q2)_{[i,j](s)}-\alpha(q2)_{[i,j](s)} < 3.20-1.60] = P[\epsilon(q2)_{[i,j](s)} < 3.20-1.60].\\
\end{eqnarray*}

McKelvey and Palfrey assume that the errors have a largest extreme value (lev) distribution and that $\alpha(q2)_{[i,j](s)}$ and $\gamma(q2)_{[i,j](s)}$ are independent leading to their subtraction $\epsilon(q2)_{[i,j](s)}$ being a logistic distribution --- the QRE Multinomial Logit model \citep{McFad73,McKPal95,McKPal96,McKPal98}.  We will follow this convention throughout the rest of the paper for computational convenience.  Based upon this distributional choice for the deviations, $q2_{[i,j]}(s)$ can be determined explicitly --- assuming the  $\epsilon$'s are distributed from the logistic distribution with precision $\lambda$ we have:

\begin{equation*}
\epsilon(q2)_{[i,j](s)} \sim \text{logistic}(shape=0, precision=\lambda),$$
$$q2_{[i,j](s)} = \frac{1}{1 + e^{-\lambda(3.20 - 1.60)}}.
\end{equation*}

\citet{McKPal98} analyze the centipede game depicted in Figure \ref{gt4low} as a game of perfect information, thus the decision probabilities for the QRE are determined via backwards induction. Using an expected utility argument, $p2_{[i,j](s)}$ can be determined from $q2_{[i,j](s)}$ and the errors in a Players A's decision. 

\begin{eqnarray*}
p2_{[i,j](s)} &=& P[U_A(y_{[i,j](s)}=3) + \alpha(p2)_{[i,j](s)} > U_A(y_{[i,j](s)}=4) + \gamma(p2)_{[i,j](s)}]\\
&=& P[U_A(y_{[i,j](s)}=3) + \alpha(p2)_{[i,j](s)} > q2 \times U_A(y_{[i,j](s)}=4) \\
&& + (1-q2_{[i,j](s)})\times U_1(y_{[i,j](s)}=5) + \gamma(p2)_{[i,j](s)}]\\
&=& P[1.60 + \alpha(p2)_{[i,j](s)} > q2_{[i,j](s)} \times 0.80 + (1-q2_{[i,j](s)}) \times 6.40 + \alpha(p2)_{[i,j](s)}]\\
&=& P[\epsilon(p2)_{[i,j](s)} < 1.60 - q2_{[i,j](s)} \times 0.80 - (1-q2_{[i,j](s)}) \times 6.40]\\
&=& \frac{1}{1 + e^{-\lambda(1.60 - q2_{[i,j](s)} \times 0.80 - (1-q2_{[i,j](s)}) \times 6.40)}}.\\
\end{eqnarray*}

By continuing to work backwards, the other two probabilities $(q1, p1)_{[i,j](s)}$ can be determined.  Based upon the four decision probabilities the probabilities of the five outcomes of the game can easily be determined as follows: 

\begin{eqnarray*}
P[y_{[i,j](s)}=1] &=& \theta^{(1)}_{[i,j](s)} = p1_{[i,j](s)}, \\
P[y_{[i,j](s)}=2] &=& \theta^{(2)}_{[i,j](s)} = (1-p1_{[i,j](s)})q1_{[i,j](s)}, \\
P[y_{[i,j](s)}=3] &=& \theta^{(3)}_{[i,j](s)} = (1-p1_{[i,j](s)})(1-q1_{[i,j](s)})p2_{[i,j](s)}, \\
P[y_{[i,j](s)}=4] &=& \theta^{(4)}_{[i,j](s)} = (1-p1_{[i,j](s)})(1-q1_{[i,j](s)})(1-p2_{[i,j](s)})q1_{[i,j](s)}, \\
P[y_{[i,j](s)}=5] &=& \theta^{(5)}_{[i,j](s)} = (1-p1_{[i,j](s)})(1-q1_{[i,j](s)})(1-p2_{[i,j](s)})(1-q1_{[i,j](s)}).
\end{eqnarray*}

\begin{figure}[htb]
\begin{center}
{\scalebox{0.6}{\includegraphics{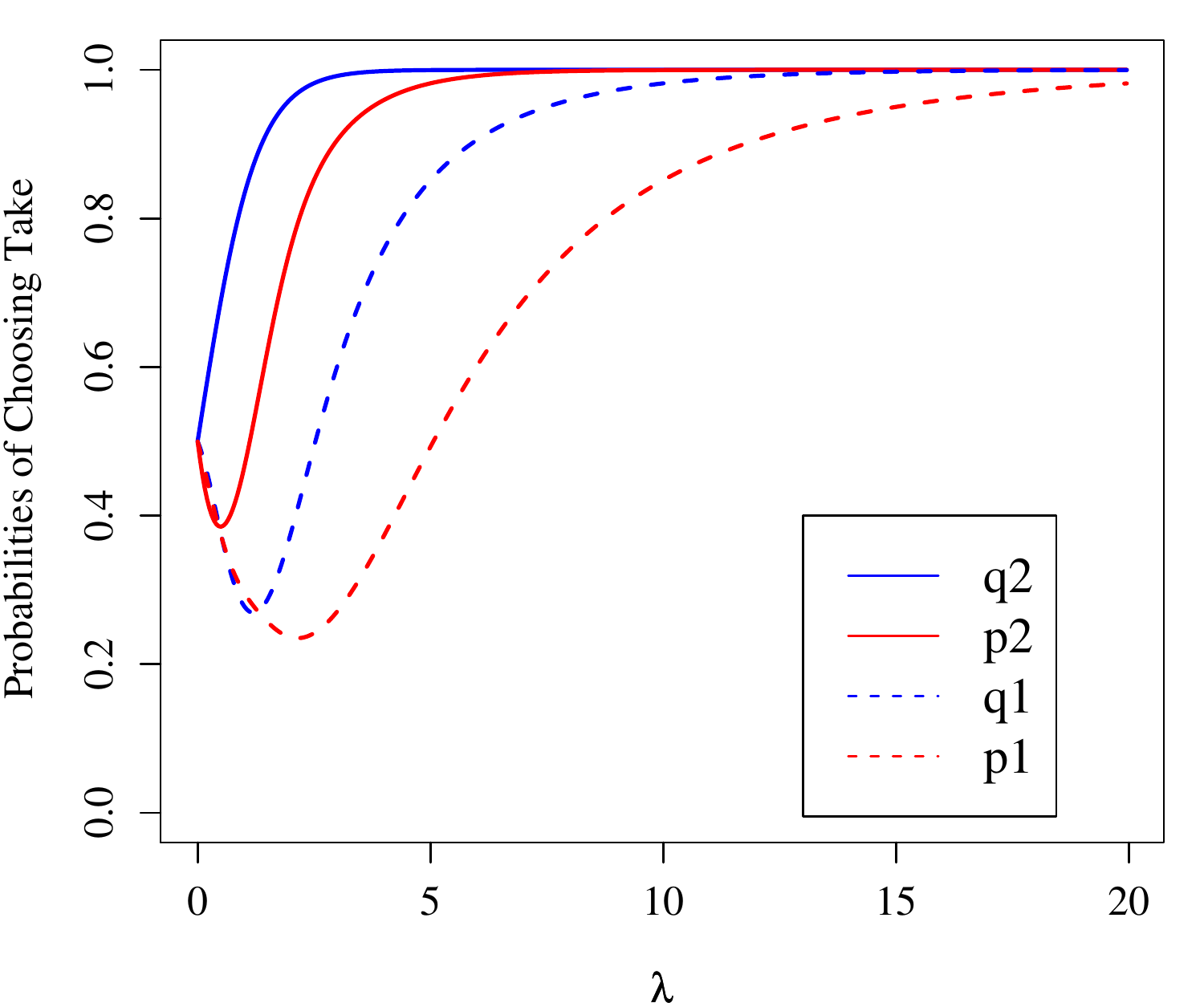}}} 
\caption{The probability of choosing Take at any point in the game goes toward 1 as the precision parameter $\lambda$ increases.}
\label{actLambda}
\end{center}
\end{figure}

Figure \ref{actLambda} shows the four different decision probabilities $(q2, p2, q1, p1)_{[i,j](s)}$ plotted as a function of $\lambda$ and in each of the four cases, as $\lambda$ increases toward $\infty$, the probability of Take  goes to 1 which is the SPNE.  When $\lambda = 0$, the probability between Take and Pass is 50/50 since a player is completely uncertain about which of the two choices is best.

\subsection{A QRE Model of Learning}
In order to examine the possibility of learning within the QRE framework, we account for the information about the $t^{th}$ game played by a pair $A_i$ and $B_j$ through the addition of a covariate to McKelvey and Palfrey's base QRE model presented in Section 4.1.  This is in line with the work of \citet{Sig99} where additional information about a game or the subjects involved allows researchers to gain an understanding of how variation in covariates leads to variation in outcomes of the game.  Figures \ref{edaHistScatt} and \ref{MCpv1} suggested a decrease in the outcome of the game as the number of games played by an individual increased.  We consider modeling this by allowing the magnitude of the precision parameter to change over repeated play --- leading to the following QRE parameterization:

\begin{eqnarray*}
(\epsilon(p1),\epsilon(q1), \epsilon(p2),\epsilon(q2))_{[i,j](s)} &\sim& \text{logistic} (shape=0, precision=\lambda e^{\beta t}).
\label{learn1}
\end{eqnarray*}

In \citet{McKPal98}, the authors had hoped that modeling heterogeneity, in terms of player type (A or B), would lead to a significant improvement in the fit of their models but they did not explore this possibility.  With this consideration, we expand our model by allowing for such differences:

\begin{eqnarray*}
(\epsilon(p1),\epsilon(q1), \epsilon(p2),\epsilon(q2))_{[i,j](s)} &\sim& \text{logistic} (shape=0, precision=\lambda_{p} e^{\beta t}), \\
\lefteqn{p=\text{player type }\in\{A,B\}.}
\label{learn2}
\end{eqnarray*}

This leads to the complete statistical specification of the model as follows:

\begin{equation}
y_{[i,j](s)} \sim \text{multinomial}((\theta^{(1)},\ldots,\theta^{(5)})_{[i,j](s)}),$$
$$(\theta^{(1)},\ldots,\theta^{(5)})_{[i,j](s)} \text{ are determined by the game tree in Figure \ref{gt4low} and}$$
$$\text{the following QRE specification:}$$
$$(\epsilon(p1),\epsilon(q1), \epsilon(p2),\epsilon(q2))_{[i,j](s)} \sim \text{logistic} (shape=0, precision=\lambda_{p} e^{\beta t}),$$
$$p=\text{player type }\in\{A,B\}.
\label{learn2}
\end{equation}

Based upon the QRE model the likelihood of the observed data is:

\begin{eqnarray*}
L(\lambda_A,\lambda_B, \beta | \bs{y}) &=& \prod_{i=1}^{N(s)}\prod_{j=1}^{N(s)} \prod_{s=1}^{3} \theta_{[i,j](s)}^{(1)^{ I\{y_{[i,j](s)}=1\}}}\times ... \times  \theta_{[i,j](s)}^{(5)^{I\{y_{[i,j](s)}= 5\}}}. \\
\end{eqnarray*}

The first two models in Table \ref{table1} are the results from fitting the two learning models --- one with a common $\lambda$ and the other with heterogeneity among player types.  The next two models (3 and 4) are for comparison and are discussed in \citet{McKPal93,McKPal98}.  The models consist of the one-parameter QRE model and a two parameter model which assumes ``that there is some small probability that players are `altruistic' (and hence choose [Pass] at every opportunity)''.  Finally, we also fit a standard ordered multinomial probit model which does not consider the underlying decision making process.  All the models were fit by maximum likelihood estimation for an expeditious estimation of the models and model comparisons via BIC.  Using the BIC as a measure of fit, Model 2 appears to represent the data better than the other models.  For this reason it was investigated further using a Bayesian approach in order to obtain credible intervals and examine goodness-of-fit statistics through the use of the posterior predictive tests.  Additionally, since this model will be further expanded by utilizing random effects, the move to Bayesian inference at this point is natural.  

\smallskip
\begin{table}[!h]
\begin{center}
\begin{footnotesize}
\begin{tabular}{|r|c||c||c||c|} \hline
 Number & Model   & Parameters & $LL^*$ & BIC  \\ \hline \hline
1 &  Slope in the variance &  $\lambda, \beta$ & -417.81 & -846.898 \\
2 & {\bf Slope in the variance with heterogeneity} &
$\lambda_A,\lambda_B,\beta$ & -380.195 & {\bf -777.31*}  \\\hline
3 & M \& P's original model       & $\lambda$ & -424.91 & -855.47  \\
4 &  M \& P's altruistic model &  $\lambda, q$ & -402.5 & -782.55\\ \hline
5 & Ordered Multinomial Probit & $\alpha_1,\alpha_2,\alpha_3,\alpha_4,\beta$ &  -376.928 & -782.048 \\  \hline
\end{tabular}
\end{footnotesize}
\caption{The table presents 5 different models that were initially investigated.  Based on the BIC,  Model 2 was investigated further.} 
\label{table1}
\end{center}
\end{table}

The Bayesian analysis for Model 2 based on Equations \ref{learn2} was conducted with the following diffuse priors:

\begin{equation*}
log(\lambda_A), log(\lambda_B) \sim \text{normal}(mean=0, variance=100),$$
$$\beta \sim \text{normal}(mean=0, variance=100).
\end{equation*}

The resulting posterior distribution is:

\begin{equation*}
\pi(\lambda_A,\lambda_B, \beta |\bs{y}) \propto L(\lambda_A,
\lambda_B,\beta|\bs{y})\times P(\lambda_A) \times P(\lambda_B)
\times P(\beta).
\end{equation*}

The Bayesian estimation for the model was conducted using the Metropolis algorithm. Each of the three parameters were updated separately.  A total of 500,000 iterations of the Metropolis algorithm were conducted and the first 20,000 were removed for burn-in.  The remaining iterations were thinned by sampling every 25th iteration resulting in 20,000 sampled values of $\lambda_A, \lambda_B,$ and $\beta$ from the posterior distribution --- diagnostics suggested convergence to the posterior distribution.

\begin{figure}[htb]
\begin{center}
{\scalebox{0.7}{\includegraphics{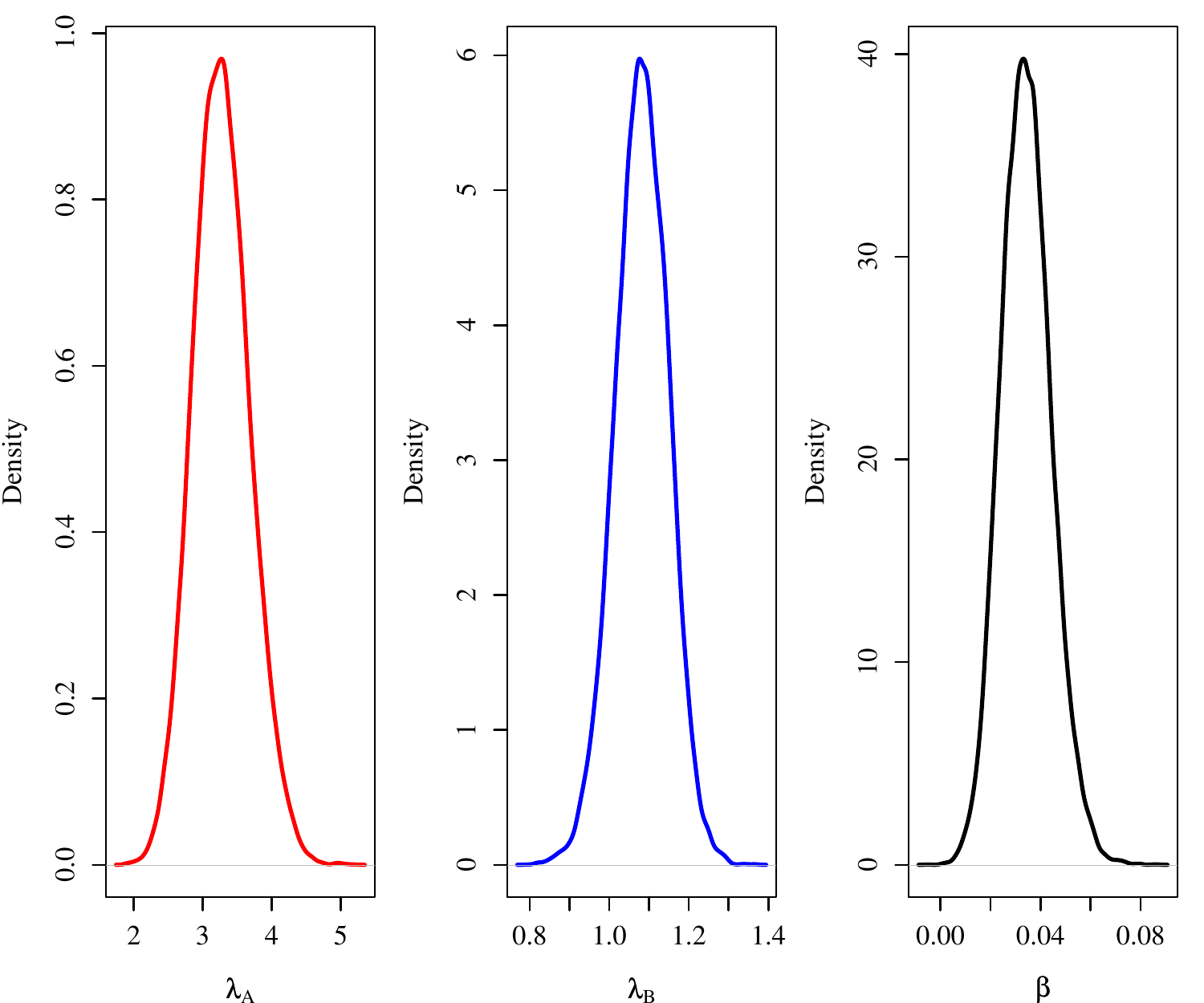}}}
\caption{The marginal posterior distributions for $\lambda_A, \lambda_B$, and $\beta$.  The posterior means are 3.275, 1.082, and 0.034.} 
\label{post1}
\end{center}
\end{figure}

Figure \ref{post1} present the densities of the posterior distributions for the three model parameters.  The main scientific question of interest depends upon the marginal posterior distribution for $\beta$.  Since the empirical $P(\beta>0)=1$, as the number of games played increases so does the precision.  We are interpreting this increase in the precision (or decrease in variance) as \textit{statistical learning}, which can also be considered as learning in the game theoretic sense since increasing the precision leads to the SPNE for the QRE model specified by Equations \ref{learn2}.  Also, it can be seen in the figure that $\lambda_B < \lambda_A$, in fact the empirical $P(\lambda_B < \lambda_A)=1$, suggesting that the Players A have higher base precision.  It is important to note that $\lambda_A$ and $\lambda_B$ not only represent each player's base precision about their utilities, but also represent each player's estimates about the precision of the other type of player.  Thus, $\lambda_B < \lambda_A$ suggests that both populations of A's and B's ``estimate'' that Players A are more ``certain'' (i.e. have a higher base level of precision) in their choices through out the game.  This matter will be discussed further in Section 5.

\begin{figure}[htb]
\begin{center}
{\scalebox{0.7}{\includegraphics{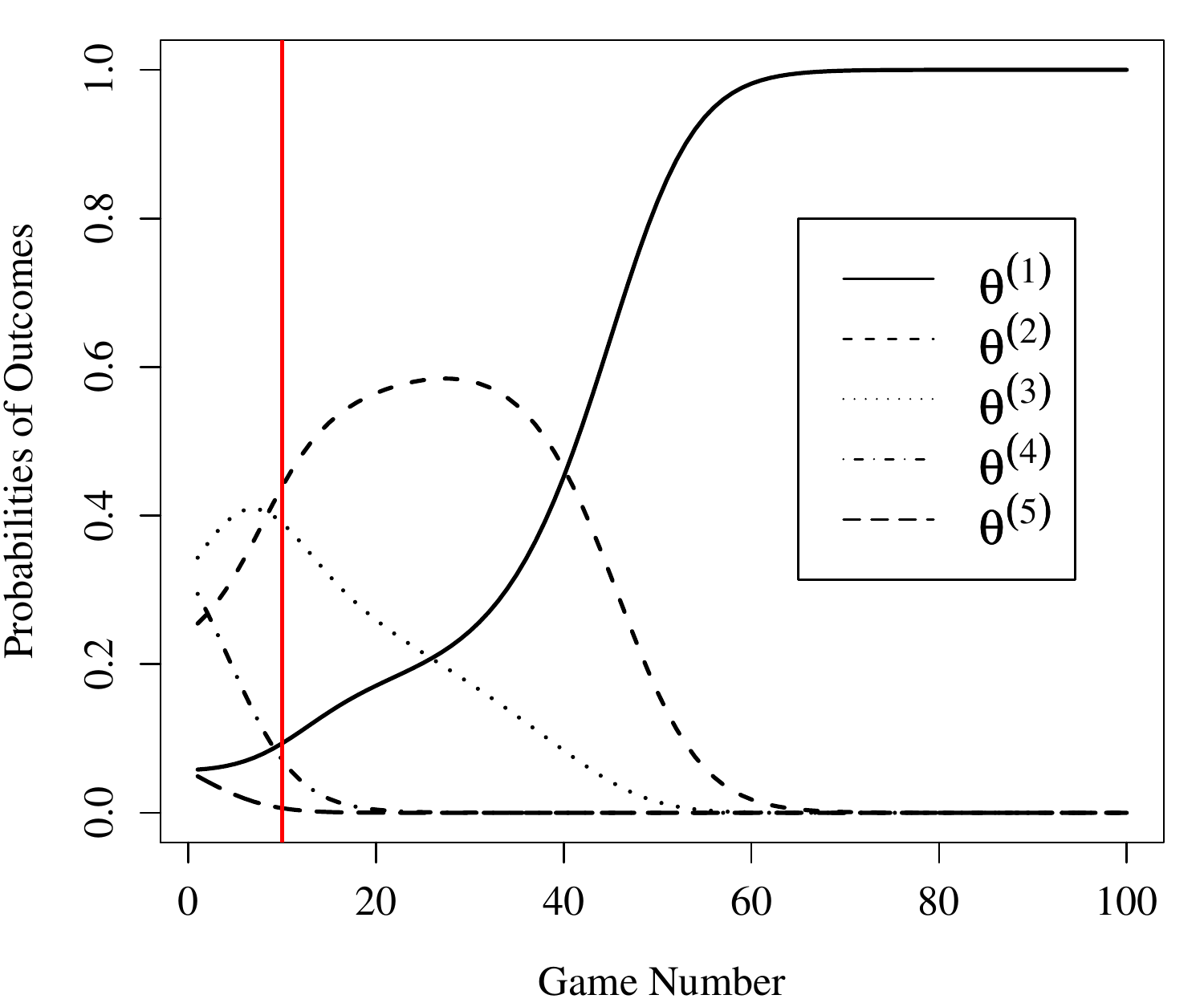}}}
\caption{Probability of the 5 outcomes vs the game number based upon the mean of the posterior distribution of the parameters from the QRE model.  The vertical line represents the end of the observed data.}
\label{probplot}
\end{center}
\end{figure}

The notion of learning can be seen more explicitly in Figure \ref{probplot}, which plots the probability of the 5 different outcomes in relation to the game number (based upon the means of the posterior distributions of the parameters).  As the number of games is extrapolated to 100, the probability of the first outcome $P(y=1)$ goes to 1.

In order to check the model fit, we conducted four posterior predictive tests \citep{GelCar}.  Each test consists of generating the posterior distribution of a test statistic of interest $T(\bs{y}_{rep}| \Delta)$ based on replicate data sets generated from the model $\bs{y}_{rep}|\Delta$ and then comparing that distribution to the test statistic computed from the observed data $T(\bs{y}_{obs})$.  Here, $\Delta$ represents the posteriors of the model parameters.  In particular, 20,000 replicate data sets $\bs{y}_{rep}$ were generated from the joint posterior distribution of the parameters based on the 20,000 MCMC scans. These tests are a way to see if the model is capturing features of interest in the observed data.  A way to quantify the notion of ``capturing'' is through Bayesian p-values based on a particular test statistic: $P[|T(\bs{y}_{rep}| \Delta)| \geq T(\bs{y}_{obs})]$.  A small p-value suggests that the model is not capturing the statistic of interest.  For this data, we are concerned with examining whether the model captures: 

\begin{enumerate}
\item the trend of the outcomes against the game numbers;
\item potential differences among the Players A;
\item potential differences between the Players B;
\item potential differences between the sessions.
\end{enumerate}

For (1), in order to make a comparison to the randomization test conducted in Section 3.1, the test statistic used was the slope of the linear trend of the outcomes versus the number of games played.  For (2)-(4), the variance between the Players A, Players B, or sessions was compared to the variance within each of these groups via an F-statistic.  Here we use the term `potential differences' since our model did not account for differences in each of those groups, which a priori may be an adequate assumption.  Before we state the four test statistics explicitly some additional notation is necessary:

\begin{displaymath}
\begin{array}{ccc}
y_{[\cdot,j](s)} = \sum_{i=1}^{N(s)} y_{[i,j](s)}, & \bar{y}_{[\cdot,j](s)} = \frac{\sum_{i=1}^{N(s)} y_{[i,j](s)}}{N(s)},  & \bar{y}_{[\cdot,j](\cdot)} = \frac{\sum_{i=1}^{N(s)} \sum_{s=1}^3{y_{[i,j](s)}}}{\sum_{s=1}^3N(s)}, \\
\\
y_{[\cdot,\cdot](s)} = \sum_{i=1}^{N(s)} \sum_{j=1}^{N(s)} y_{[i,j](s)}, & \bar{y}_{[\cdot,\cdot](s)} = \frac{\sum_{i=1}^{N(s)} \sum_{j=1}^{N(s)}  y_{[i,j](s)}}{N(s)\times N(s)},  & \bar{y}_{[\cdot,\cdot](\cdot)} = \frac{\sum_{i=1}^{N(s)} \sum_{j=1}^{N(s)} \sum_{s=1}^3{y_{[i,j](s)}}}{\sum_{s=1}^3N(s)\times N(s)}. \\

\end{array}
\end{displaymath} 

\noindent In particular the four test statistics we considered are:

\begin{enumerate}
\item $\gamma = \frac{\sum_{i=1}^{N(s)}\sum_{j=1}^{N(s)} \sum_{s=1}^{3} [(t_{[i,j](s)} -  \bar{t}_{[\cdot,\cdot](\cdot)})(y_{[i,j](s)} -  \bar{y}_{[\cdot,\cdot](\cdot)})]}{\sum_{i=1}^{N(s)}\sum_{j=1}^{N(s)} \sum_{s=1}^{3} (t_{[i,j](s)} -  \bar{t}_{[\cdot,\cdot](\cdot)})^2}$;

\item $F_{\textrm{Players A}} = \frac{\sum_{i=1}^{N(s)}\sum_{j=1}^{N(s)} \sum_{s=1}^{3}(\bar{y}_{[i,\cdot](\cdot)} -  \bar{y}_{[\cdot,\cdot](\cdot)})^2/(\sum_{s=1}^3N(s)-1)}{\sum_{i=1}^{N(s)}\sum_{j=1}^{N(s)} \sum_{s=1}^{3}(y_{[i,j](s)} -  \bar{y}_{[i,\cdot](\cdot)})^2/(\sum_{s=1}^3N(s)\times N(s)-\sum_{s=1}^3N(s))} = \frac{MS_{\textrm{Players A}}}{MS_{\textrm{error}}}$;

\item $F_{\textrm{Players B}} = \frac{\sum_{i=1}^{N(s)}\sum_{j=1}^{N(s)} \sum_{s=1}^{3}(\bar{y}_{[\cdot,j](\cdot)} -  \bar{y}_{[\cdot,\cdot](\cdot)})^2/(\sum_{s=1}^3N(s)-1)}{\sum_{i=1}^{N(s)}\sum_{j=1}^{N(s)} \sum_{s=1}^{3}(y_{[i,j](s)} -  \bar{y}_{[\cdot,j](\cdot)})^2/(\sum_{s=1}^3N(s)\times N(s)-\sum_{s=1}^3N(s)))} = \frac{MS_{\textrm{Players B}}}{MS_{\textrm{error}}}$;

\item $F_{\textrm{sessions}} = \frac{\sum_{i=1}^{N(s)}\sum_{j=1}^{N(s)} \sum_{s=1}^{3}(\bar{y}_{[\cdot,\cdot](s)} -  \bar{y}_{[\cdot,\cdot](\cdot)})^2/(3-1)}{\sum_{i=1}^{N(s)}\sum_{j=1}^{N(s)} \sum_{s=1}^{3}(y_{[i,j](s)} -  \bar{y}_{[\cdot,\cdot](s)})^2/(\sum_{s=1}^3N(s)\times N(s) -3)} = \frac{MS_{\textrm{sessions}}}{MS_{\textrm{error}}}$.
\end{enumerate}

\begin{figure}[htb]
\begin{center}
{\scalebox{0.9}{\includegraphics{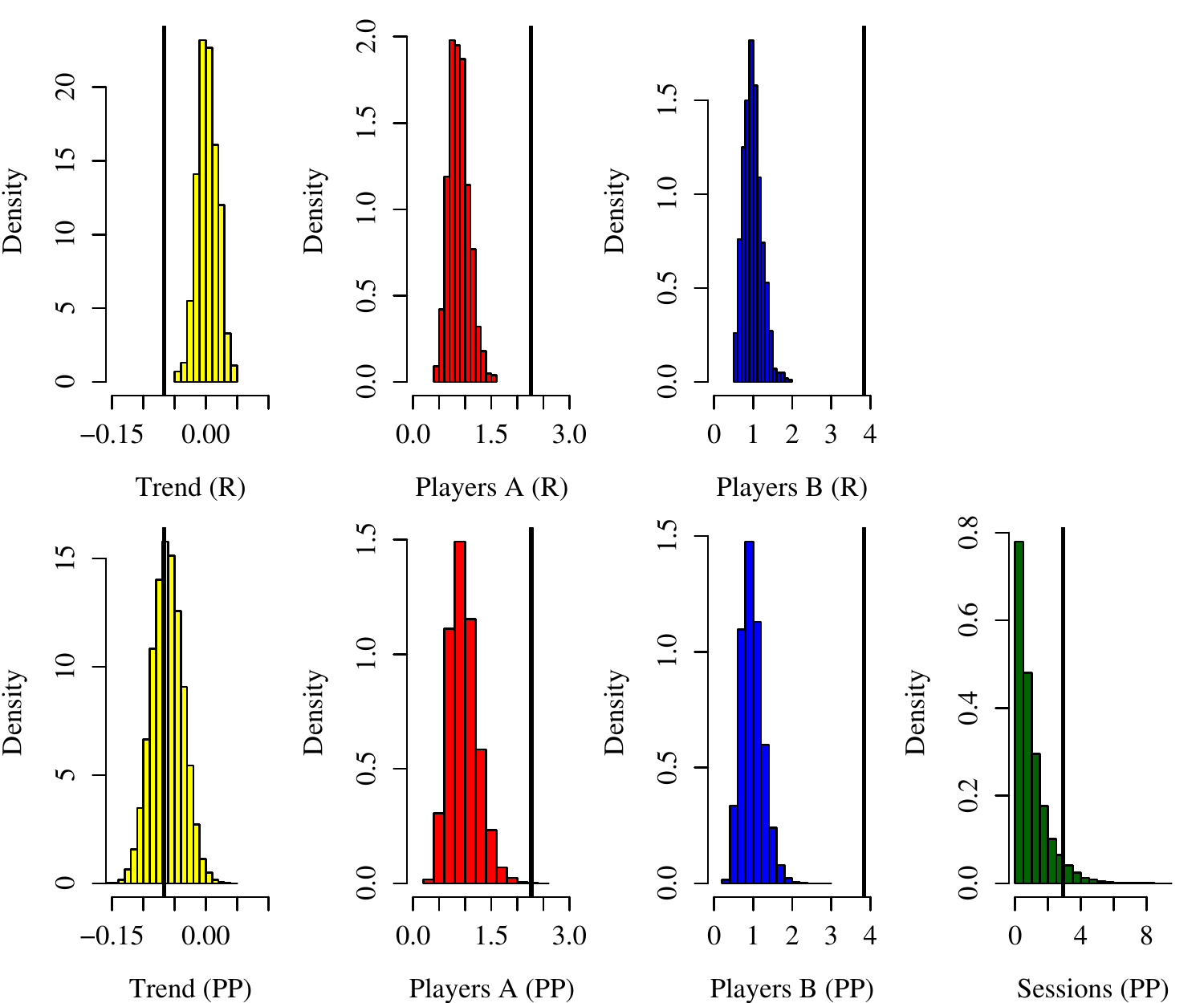}}}
\caption{The top row are results from the randomization tests (R) and the bottom row, are results from the posterior predictive tests (PP).  In each column the same test statistics were used.}
\label{3ParamTests}
\end{center}
\end{figure}

The second row of Figure \ref{3ParamTests} presents the results for the posterior predictive tests (PP).  The first panel in that row depicts the posterior distribution of the slope $\gamma(\bs{y}_{rep}|\Delta)$ while the vertical line represents the test statistic from the observed data $\gamma(\bs{y}_{obs})$.  The one sided Bayesian p-value for this posterior predictive test is $P[\gamma(\bs{y}_{rep} |\Delta) \leq \gamma(\bs{y}_{obs})] = 0.421$, which suggests that model is capturing the slope for the linear trend.  In comparison, the Bayesian p-values for the tests examining the differences among the Players A and Players B are both zero, suggesting that the model could be expanded to allow for differences among the subjects within each player type.  Finally, the last panel suggests that incorporating a session effect into the model may not be necessary since $P[F_{\textrm{sessions}}(\bs{y}_{rep}| \Delta) \geq F_{\textrm{sessions}}(\bs{y}_{obs})] = 0.056$, considering 0.05 a cut-off value.

In comparison, the top row of Figure \ref{3ParamTests} presents the results for a set of randomization tests.  The first panel replicates the results from Section 3.1 and allows for a comparison between the randomization test and the posterior predictive test when examining the trend.   Recall that the hypothesis for the randomization test was $H_0$: the game number does not affect the outcome.  From this randomization test, we concluded that we could reject the null hypothesis.  Now through modeling the trend, the posterior regard test suggests that our model is capturing that trend.  This approach allows one to initially investigate a set of hypotheses of interest via the randomization testing approach and then using the same test statistics compare the results based upon a particular model using posterior predictive tests.   In regard to randomization tests and hypothesis testing in general \citeposs{Bes77} state ``we contend that significance testing is rarely to be treated as an end in itself, its purpose being more usually as an aid in suggesting further hypotheses relevant data collection''.  Or in this case, as a preliminary tool for exploratory data analysis which can lead to further modeling.  The next two panels in the top row examine the following two null hypothesis:

\begin{enumerate}
\item $H_0$: there are no differences among the Players A;
\item $H_0$: there are no differences among the Players B.
\end{enumerate}

The following procedure was used for the randomization tests: 1.) within each session, a random permutation of the Latin square, with either Players A in the center or Players B in the table entries, was conducted under the null hypotheses of no difference; 2.) the three sessions were combined and the appropriate F-statistic was computed. 3.) This procedure was repeated 1,000 times leading to the null distributions displayed in Figure \ref{3ParamTests}.  From the histograms labeled `Players A (R)' and `Players B (R)' it is clear that the p-values are zero, and we should reject the null hypotheses.  These results coincide with the posterior predictive tests, which were based on a model which did not account for differences among the subjects.  Finally, we are unable to conduct a randomization test for differences among the three sessions which is faithful to the design.   Since different subjects are nested within each session and each session has Latin square design, we cannot simply permute a subject between sessions without destroying the Latin square design.  We could conduct a randomization test which ignored the design, these are typically called `unrestricted' randomization tests since each subject can be allocated to any treatment combination however as \citet{GarJolJon95} note ``[t]his has disadvantages for testing whether one factor affected the responses, since the influence of other factors may bias the results''.  Since we have a model at this point, we will rely on a posterior predictive test to examine the question of differences among sessions and forgo the unrestricted randomization test.

\section{A QRE Random Effects Model of Learning}
All of the previous models have assumed no subject specific effect for the Players A or the Players B;  however, repeated observations from the same subject could result in statistical dependence of the outcomes which are assumed to be independent in the previous QRE models.  As was noted from both the randomization tests as well as the posterior predictive tests there does appear to be substantial differences between the subjects (either Players A or Players B).  In order to account for this correlation, a random effects model was employed.  A QRE random effects model is easily developed through a Bayesian hierarchical approach. Now, each subject has their own set of parameters which come from a population of parameters  for each player types --- $\lambda_{A_i}, \beta_{A_i}, \lambda_{B_j}$, and $\beta_{B_j}$.

\begin{figure}[htb]
\begin{center}
{\scalebox{0.7}{\includegraphics{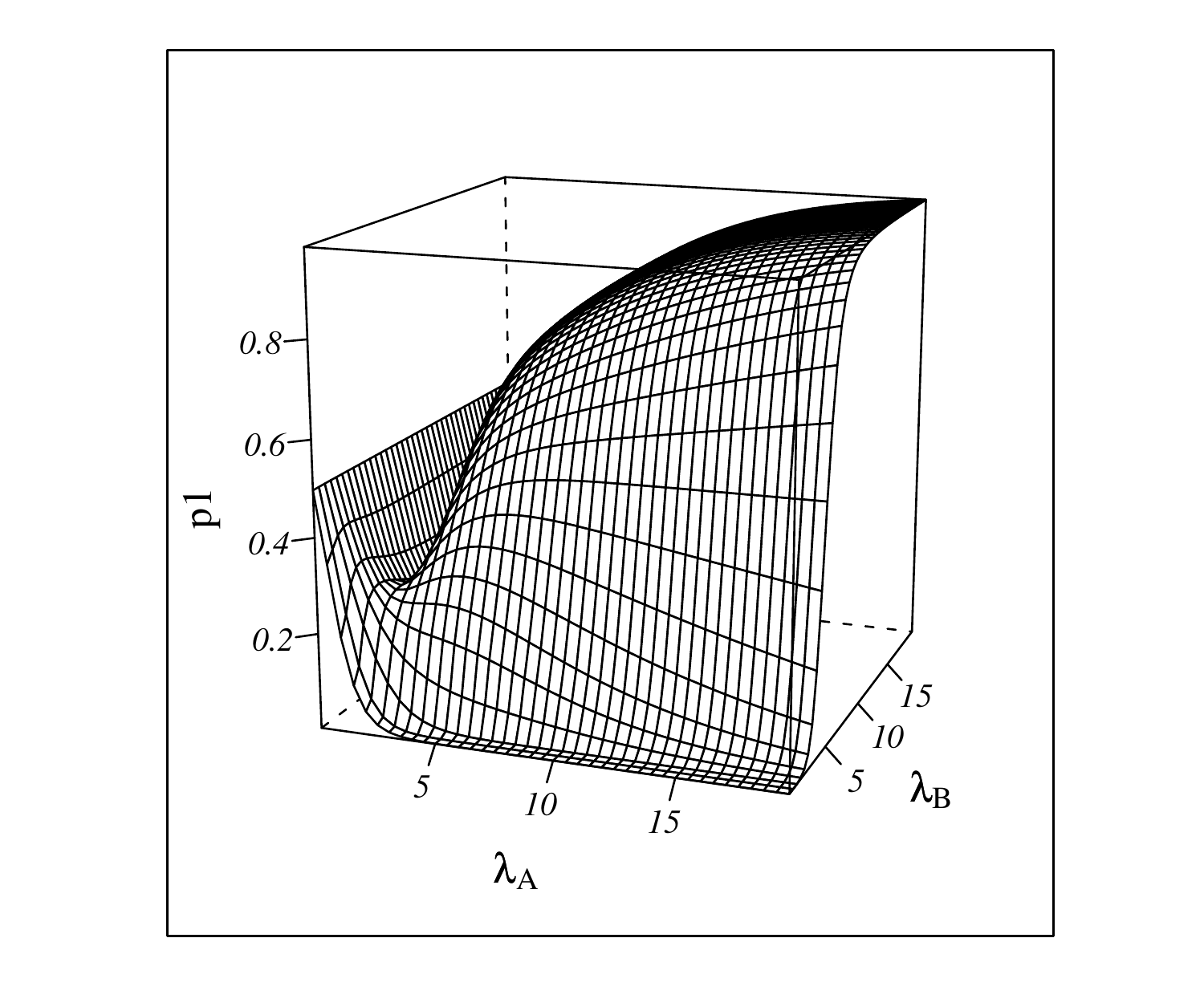}}} 
\caption{The figure shows that the probability that for a Player A to choose take at the first stage of the game (p1) depends not only on Player A's precision parameter $\lambda_A$ but also on the precision parameter $\lambda_B$ of Player B.}
\label{persp}
\end{center}
\end{figure}

Another important consideration is that the experimental design is such that players do not know whom they are playing against.  The QRE however assumes that every player knows every other player's error distribution.  In the simple case, where we modeled two player types (Players A and Players B) and did not consider individual subject effects.  This probabilistic model induces relationships between $\lambda_A,\lambda_B$ and the probabilities of choosing Take at various stages of the game $(q2,p2,q1,p1)_{[i,j](s)}$ that are not as simple as in the one parameter model depicted in Figure \ref{actLambda}.  To clarify the point, Figure \ref{persp} demonstrates the relationship between $\lambda_A$, $\lambda_B$, and $p1$ (probability that Player A will choose Take at the first stage of the game), based on the model defined by Equations \ref{learn2}, where with out loss of generality, we take $\beta=0$. The figure shows that simple statements can not be made about $p1$ as $\lambda_A$ increases.  The probability for Player A to choose Take at stage 1 depends upon the specification of $\lambda_B$, and could go to 1 or to 0 (away from the SPNE) as $\lambda_A$ increases.  The reason for this lies in the assumptions about the QRE model, in that the players' are assumed to know everyone else's error distribution.  Thus, if $\lambda_B$ is small, suggesting that Player B is equally indifferent between choosing Take or Pass, then Player A will maximize her expected utility by choosing Pass as $\lambda_A$ increases. This is the point that \citet{McKPal96} make in their example about chess players: ``[Consider] a chess game between an expert and a beginner.  If this were common knowledge, then the expert might adopt a different strategy than she would against another expert.''

When we only model player types, the assumption that that subjects know the parameters may not seem unreasonable, but when we move to a random effects model where each subject has there own set of parameters, then this assumption of knowledge about the other players is far too strong.  This is especially important considering that the subjects did not know whom they were playing against.  To return to a slightly less restrictive assumption, we assume that subjects may not know the distribution of the subjects they are playing against but the empirical means of parameters associated with the opposing player type. Thus each choice is a probabilistic function of subject specific $\lambda$ and $\beta$ parameters, empirical means of the $\lambda$ and $\beta$ parameters of the opposing player type, and the current game number:

\begin{equation}
\text{Player As' decisions:}$$
$$P(Take)_{A_i} =
F_{A_i}(\lambda_{A_i},\beta_{A_i}, \bar{\lambda}_B, \bar{\beta}_B, t); \ \ \ i \in \{1,\ldots,29\}.$$
$$\text{Player Bs' decisions:}$$
$$P(Take)_{B_j} =
F_{B_j}(\lambda_{B_j},\beta_{B_j}, \bar{\lambda}_A, \bar{\beta}_A, t);  \ \ \ j \in \{1,\ldots,29\}.
\label{genprob}
\end{equation}
\medskip

The above can further be clarified as follows.  The probability that Player $B_j$ will choose Take at stage 4 of the game ($q2$) when playing against Player $A_i$ is given by:

\begin{eqnarray*}
q2_{[i,j](s)} &=&  logit\left[\lambda_{B_j}exp(\beta_{B_j} t_{[i,j](s)})(3.20 - 1.60)\right].\\
\end{eqnarray*}

Now the probability that Player $A_i$ will choose Take at stage 3 of the game ($p2$) when playing against Player $B_j$ depends on empirical means of the Players B:

\begin{eqnarray*}
p2_{[i,j](s)}   &= & logit\left[\lambda_{A_i} exp(\beta_{A_i} t_{[i,j](s)})(1.60 - \bar{q2}_{[i,j](s)}0.80 - (1-\bar{q2}_{[i,j](s)})6.40)\right], \\ \\
\bar{q2}_{[i,j](s)} &=& logit\left[\bar{\lambda}_{B_j}exp(\bar{\beta}_{B_j} t_{[i,j](s)})(3.20 - 1.60)\right]. \\
\end{eqnarray*}

Next the probability that Player $B_j$ will choose Take at stage 2 of the game ($q1$) when playing against Player $A_i$ depends on the empirical means related to the Players A.  Note that in Equation 4, $q2_{[i,j](s)}$ is known to Player $B_j$ when determining $q1_{[i,j](s)}$ as long as it is not embedded in $\bar{p2}_{[i,j](s)}$ which is the case in Equation 5.  Thus subjects are always assumed to know their own probabilities as long as they are not embedded in another subject's decision probabilities.  We feel this is a reasonable assumption and is much preferred to subjects knowing exactly the parameters of their opponents, however it is an area of inquiry which could be inspected further.

\begin{eqnarray}
q1_{[i,j](s)}  &=& logit \left[\lambda_{B_j}exp(\beta_{B_j} t_{[i,j](s)})(0.80 - \bar{p2}_{[i,j](s)}0.40 - \right. \nonumber \\
&& \left. ( 1-\bar{p2}_{[i,j](s)})q2_{[i,j](s)}3.20 - (1-\bar{p2}_{[i,j](s)})(1-q2_{[i,j](s)})1.60)\right],\\  \nonumber \\
 \bar{p2}_{[i,j](s)} &=& logit \left[ \bar{\lambda}_{A_i}exp( \bar{\beta}_{A_i} t_{[i,j](s)})(1.60 - \bar{q2}_{[i,j](s)}0.80 - \right. \nonumber \\
 && \left.   (1-\bar{q2}_{[i,j](s)})6.40)\right].
\end{eqnarray}

\begin{figure}[!htb]
\begin{center}
{\scalebox{0.7}{\includegraphics{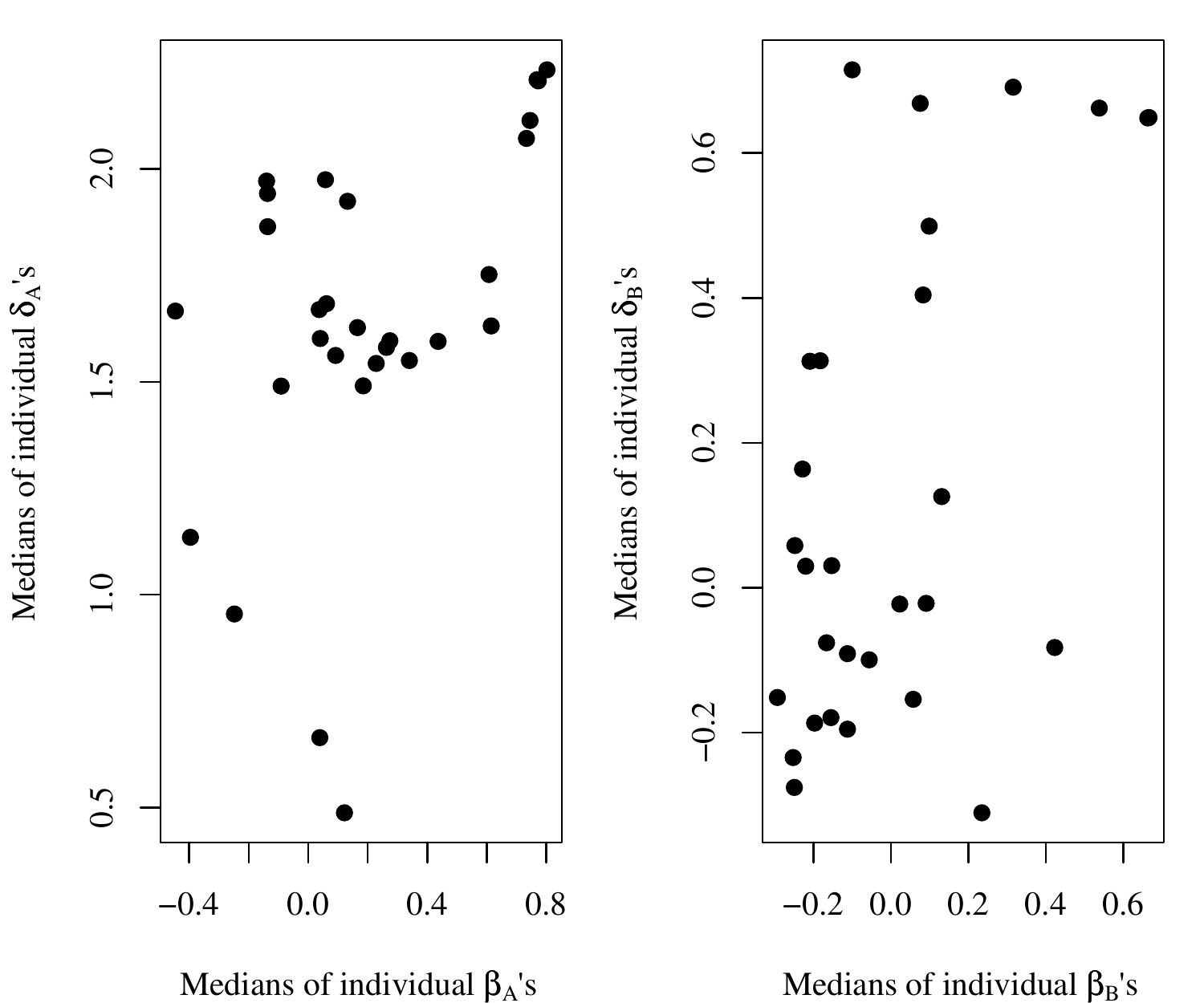}}} 
\caption{Scatter plots of the medians of each subjects slope and intercept for Players A and Players B. }
\label{deltabeta}
\end{center}
\end{figure}

Based on these considerations, we can fully define the QRE random effects model.  Here again we transform the precision terms $\lambda_{A_i}, \lambda_{B_j}$ by modeling $\delta_{A_i} = log(\lambda_{A_i}), \delta_{B_j} = log( \lambda_{B_j})$, which creates a random slope and intercept form of the model.  A priori we assume the the sampling distributions for the intercepts and slopes for each subject are not correlated, which appears to be justified by the scatter plots in Figure \ref{deltabeta}, since no strong linear pattern is present.  Finally, the priors were chosen to be conjugate but diffuse.  Altogether the model is:

\begin{equation}
y_{[i,j](s)} \sim \text{multinomial}(\theta^{(1)}_{[i,j](s)},\ldots,\theta^{(5)}_{[i,j](s)}).$$
$$\theta^{(1)}_{[i,j](s)},\ldots,\theta^{(5)}_{[i,j](s)} \text{ are determined by the game tree in Figure \ref{gt4low} and}$$
$$\text{the following QRE specifications:}$$
$$\textrm{Players A: } (\epsilon(p1), \epsilon(p2))_{[i,j](s)} \sim \text{logistic} (shape=0,
precision=\lambda_{A_i}e^{{\beta_{A_i} t}}=e^{log(\lambda_{A_i})+
{\beta_{A_i} t}}=e^{\delta_{A_i}+ {\beta_{A_i} t}}).$$
$$\textrm{Players B: } (\epsilon(q1), \epsilon(q2))_{[i,j](s)} \sim \text{logistic} (shape=0,
precision=\lambda_{B_j}e^{{\beta_{B_j} t}}=e^{log(\lambda_{B_j})+
{\beta_{B_j} t}}=e^{\delta_{B_j}+ {\beta_{B_j} t}}).$$
$$\delta_{A_i} \sim \text{normal}( mean = \mu_{\delta_A}, variance=\sigma_{\delta_A}^{2}),$$
$$\delta_{B_j} \sim \text{normal}(mean = \mu_{\delta_B}, variance=\sigma_{\delta_B}^{2}),$$
$$\beta_{A_i} \sim \text{normal}( mean = \mu_{\beta_A}, variance=\sigma_{\beta_A}^{2}),$$
$$\beta_{B_j} \sim \text{normal}(mean = \mu_{\beta_B}, variance=\sigma_{\beta_B}^{2}),$$
$$\mu_{\delta_A}, \mu_{\delta_B}, \mu_{\beta_A}, \mu_{\beta_B} \sim \text{normal}(mean=0, variance=100),$$
$$\sigma_{\delta_A}^{2}, \sigma_{\delta_B}^{2},\sigma_{\beta_A}^{2}, \sigma_{\beta_B}^{2} \sim \text{inverse-gamma} (1, 1).
\label{re}
\end{equation}

The estimation of the model parameters was through the construction of a Markov chain.  We found the mixing of the parameters related to the Players B to be very slow, so we conducted a total of 20 million scans, of which the first 5 million were dropped for burn-in. We thinned the remaining scans by taking every 1,000th, which left us with 15,000 samples from the posterior distribution.  Again the chains were inspected for convergence.

\begin{figure}[!htb]
\begin{center}
{\scalebox{0.8}{\includegraphics{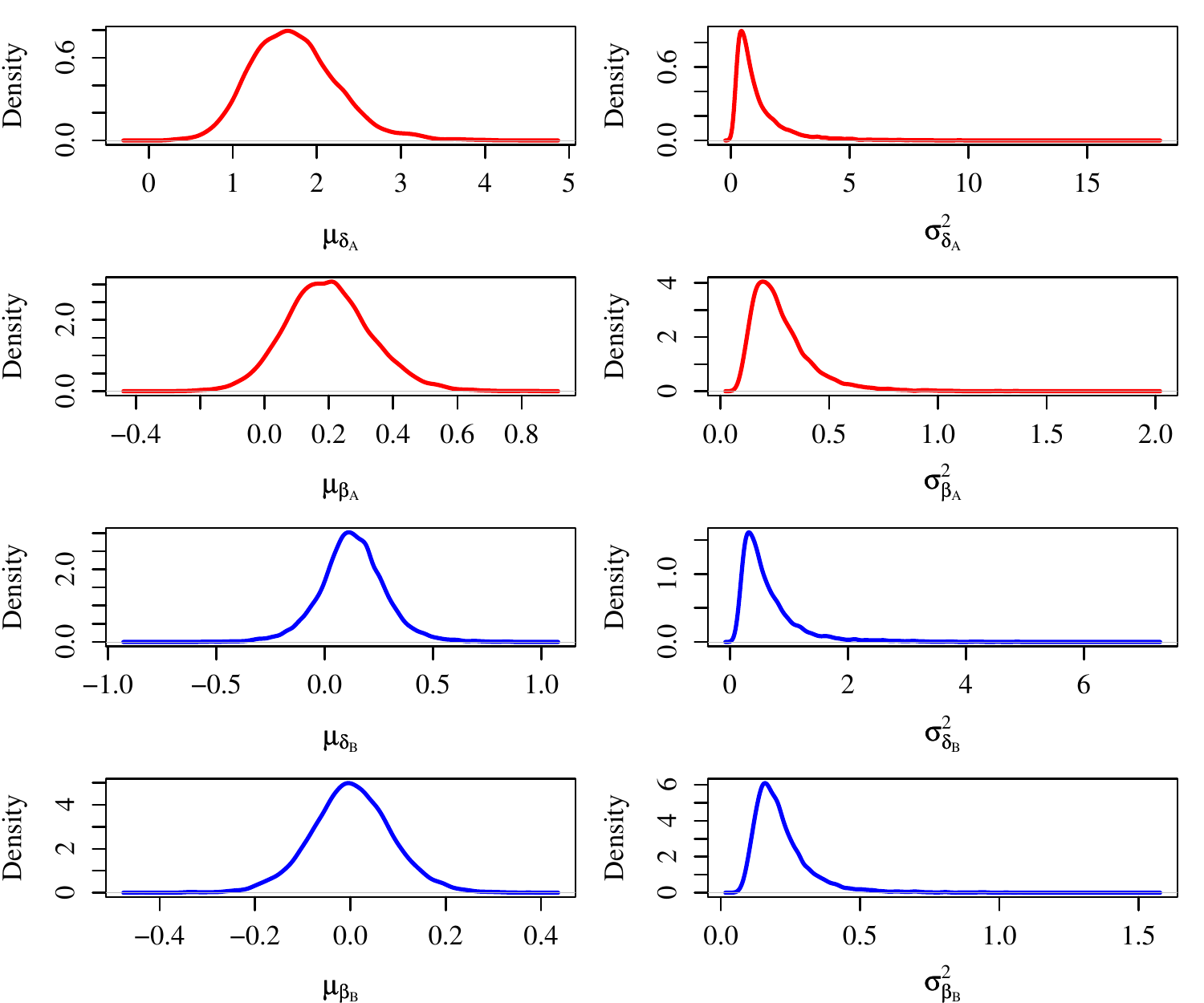}}}
\caption{Posterior distributions of population parameters.}
\label{popre}
\end{center}
\end{figure}

\begin{figure}[!htb]
\begin{center}
{\scalebox{0.7}{\includegraphics{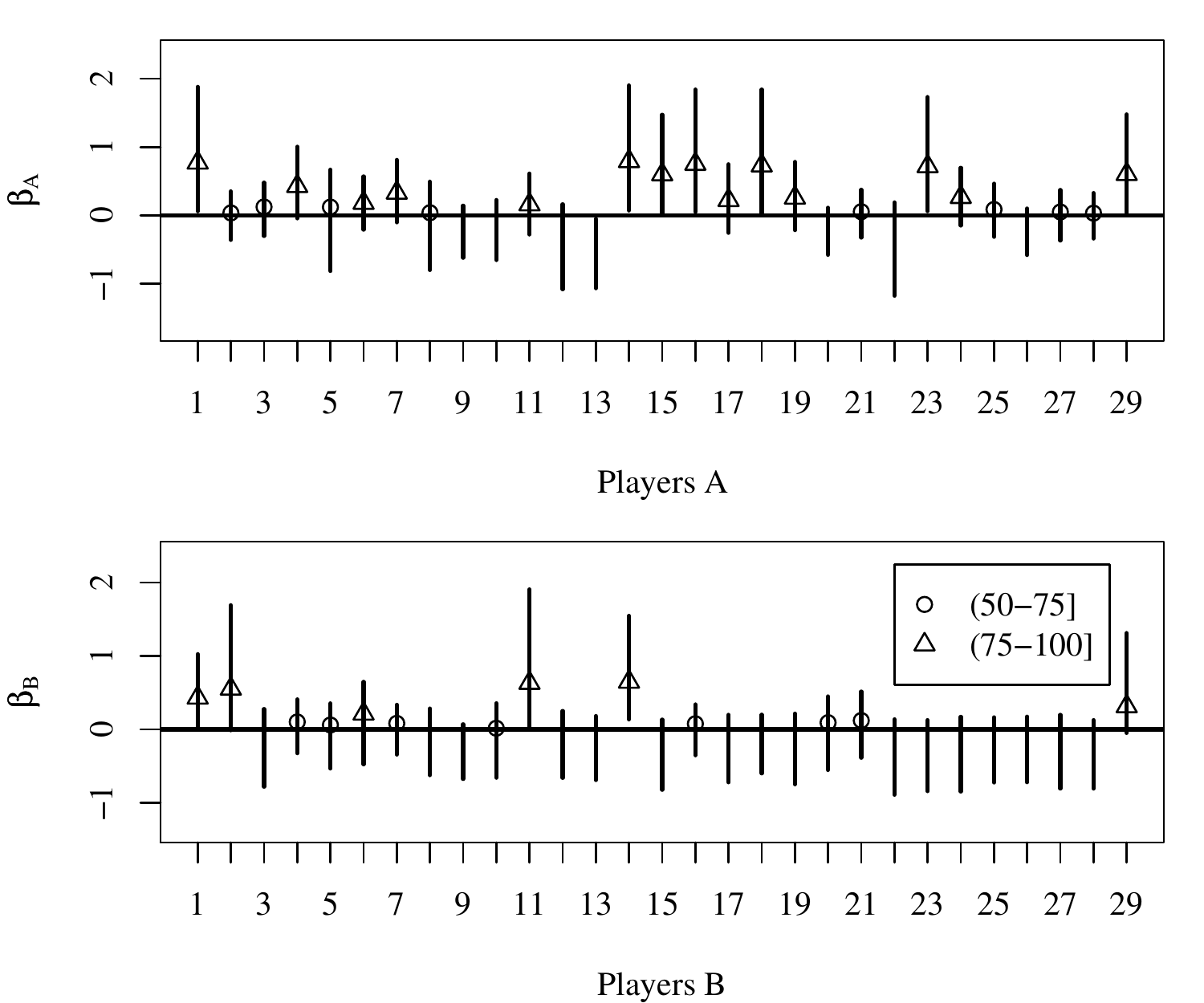}}} 
\caption{95\% credible intervals $\beta$'s for each subject.  The dots and triangles in the figure are the medians and represent the associated probability that $\beta$ for that subject is greater than zero.} 
\label{betas}
\end{center}
\end{figure}

The main parameters of interest are the population means and variances of $\delta_{A_i}, \delta_{B_j}, \beta_{A_i},$ and $\beta_{B_j}$ whose posterior distributions can be seen in Figure \ref{popre}.  The $\mu_{\delta}$'s represent a base level of precision for the population of Players A and Players B.  Again we see that, on average, the Players A have a higher base precision, but their base precision is more variable compared to the Players B.  The figure also presents the posterior distributions of the means and variances of the $\beta_{A_i}$'s and $\beta_{B_j}$'s.  The medians of the posterior distributions of  $\mu_{\beta_A}$ and  $\mu_{\beta_B}$ are greater than zero.  While the 95\% credible intervals contain zero in both cases, 95\% and 51\% of the population A and population B, respectively will have a mean for $\beta$ that is greater than zero.  Thus for that proportion of the population, as the game number increases so will the precision, which we interpret as learning in both the statistical and game-theoretic senses.  Again there does appear to be greater variation among the Players A compared to the Players B in regards to $\beta$'s, but this is not nearly as great as was seen for the $\delta$'s.  Additionally, we can also compare the distribution of $\beta$'s for each subject by examining the 95\% credible intervals in Figure \ref{betas}.  The dots and triangles in the figure are the medians.  The dots represent subjects where the probability that $\beta$ for that subject is greater than 0 is between 50\% and 75\%.  The triangles represent subjects where the probability that $\beta$ for that subject is greater than 0 is between 75\% and 100\%.  For the Players A, 22 out of 29 have medians greater than zero, compared to Players B where it is only 13 out of 29 have medians greater than zero.  Finally, the plot in Figure \ref{deltabeta} depicts a scatter plot of the medians for each subject's $\delta$ and $\beta$ grouped again by player type.  The lack of linearity, or more precisely sphericity, justifies our model assumptions of no correlation between the slope and intercept for each subject.  

\begin{figure}[!htb]
\begin{center}
{\scalebox{0.9}{\includegraphics{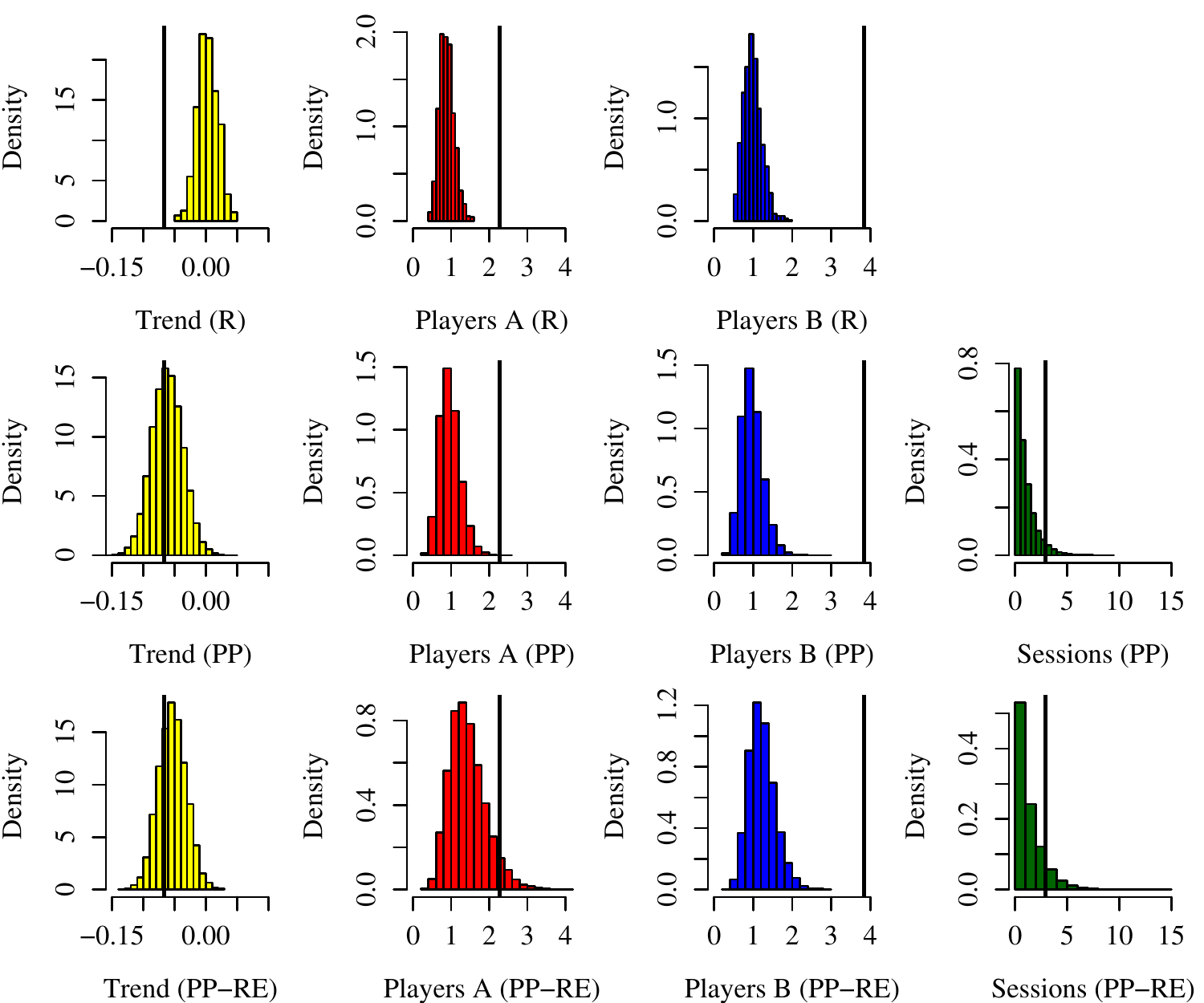}}}
\caption{The last row presents the posterior predictive tests for the random effects QRE model.  Except for the Players B, the model captures features of the observed data that we are interested in.} 
\label{ppre}
\end{center}
\end{figure}

As before, it is important to check the fit of the model compared to the data using the posterior predictive tests.  The same set of test statistics were employed and the results can be seen in Figure \ref{ppre}.  The first and second rows are the randomization tests and posterior predictive tests from Figure \ref{3ParamTests}.  The last row presents the posterior predictive tests for the random effects model (PP-RE).  The p-values associated with these tests are 0.272,  0.057, 0, and 0.118, respectively.  From these results, it appears that we are capturing the features related to the trend and differences between the Players A.  While there is an extremely slight shift in the histograms examining the differences between the Players B between the models with and without random effects, this is a feature which the model still does not appropriately capture.  Since some Players B do pass at the last stage of the game, incorporating some version of the `altruistic' model suggested by \citet{McKPal98} may lead to a better fit.  A two component mixture model for the random effects associated with the Players B may pick up on this notion of altruism.  However, with only 29 subjects the estimation may really heavily on the priors, since potentially only a few individuals would end up in the `altruistic' group.   Finally, since we allow for differences for each subject through the random effects, we would expect to be able to also capture potential differences between the sessions.

\section{Conclusion}
In this article we have examined expansions of the QRE model to allow for game-theoretic learning via repeated play of a game.  This was done by allowing the precision of players' error distributions to change as players gain experience with the game. In analyzing a data set on repeated plays of the two player centipede game, it was found that such a model fits better than the standard QRE model. The model was also expanded  to allow for heterogeneity across experimental subjects by introducing  random effects terms to capture variability in how players learn.  This requires a modification of the QRE formulation, as it is unlikely  that players know how fast their fellow game-players are  learning (at least in the data sets considered in this paper). Instead, we assume each player makes a  ``best guess'' at their fellow players behavior, based on population averages of the parameters describing behavior.  Additionally, we employed both randomization tests for exploratory data analyses and posterior predictive tests for both exploratory and confirmatory data analyses. 
	
Several extensions to the approaches in this paper can be taken.  We assumed in the random effects model that player's are always assumed to know their own probabilities as long as they are not embedded in the other player's decision probabilities, but we could globally allow them to know their own probabilities.  As mentioned, it would be interesting to explore mixture models with larger data sets or simulated data.  Finally, another issue that needs to  be addressed is the one of the  distribution of player errors.  QRE models, as well as, statistical choice models in general are constrained by the type of player error distributions that are employed, being either largest extreme value of Gaussian. \cite{QuiWes09} have utilized a semi-parametric approach to solve this problem within the QRE framework.  This method could also be applied with the QRE random effects model to allow for greater flexibility in the modeling and testing of learning --- again a larger data set would be needed so that priors are not completely dominating the results.

More generally, the modeling approach in this paper can be seen as an  attempt to inform a statistical analysis of a complicated data set with an underlying behavioral model, or conversely, expand upon a  behavioral model to allow for a more accurate description of observed data by incorporating certain characteristics of natural variability. Approaches such as this may help bridge the gap between the purely statistical analyses of social relations data \citet{Was94,GilSwa01,Hof02,Hof03,Hofbi} and game-theoretic models based on rational choice theory. 

\pagebreak
\bibliography{cent.bbl}
\end{document}